\documentclass[fleqn,usenatbib]{mnras}
\usepackage{newtxtext,newtxmath}
\usepackage[T1]{fontenc}
\DeclareRobustCommand{\VAN}[3]{#2}
\let\VANthebibliography\thebibliography
\def\thebibliography{\DeclareRobustCommand{\VAN}[3]{##3}\VANthebibliography}
\usepackage{graphicx}	
\usepackage{amsmath}	
\usepackage{amssymb}	

\title[]{Trans-Alfvenic Magnetohydrodynamic Turbulence in the Vicinity of Supernova Remnant Cassiopeia-A Shocks}
\author[P. K. Vishwakarma $\&$ J. Kumar]{
Pavan Kumar Vishwakarma$^{1}$\thanks{pavankv.rs.phy15@itbhu.ac.in}, Jais Kumar $^{1}$
\\
$^{1}$Department of Physics, Indian Institute of Technology (BHU), Varanasi, India\\
}

\begin{document}
\label{firstpage}
\pagerange{\pageref{firstpage}--\pageref{lastpage}}
\maketitle

\begin{abstract}
Statistics of the magnetic field disturbances in the supernova remnants (SNRs) can be accessed using the second-order correlation function of the synchrotron intensities. Here we measure the magnetic energy spectra in supernova remnant Cassiopeia-A by two-point correlation of the synchrotron intensities, using a recently developed unbiased method. The measured magnetic energy spectra in the vicinity of supernova remnant shocks are found to be 2/3 power law over the decade of range scales, showing the developed trans-Alfvenic magnetohydrodynamic turbulence. Our results are globally consistent with the theoretical prediction of trans-Alfvenic Mach number in developed magneto-hydrodynamic turbulence and can be explained by the amplification of the magnetic field in the vicinity of SNR shocks. The magnetic energy spectra predict SNR Cassiopeia-A having an additional subshock in the radio frequency observation along with forward and reverse shocks, with a radial window of the amplified magnetic field of $\sim$ 0.115pc near the shocks.
\end{abstract}
\section{Introduction}
\label{sec:Intro}
Supernovae and their remnants play a very crucial role in heating the interstellar medium (ISM), enriching it with heavy metal elements and accelerating the cosmic rays (CRs) within it. According to their structures supernovae remnants are mainly of 3 types: shell type, filled center type, and composite type \citep{1988ARA&A..26..295W}. Even though it is believed that supernova remnants (SNRs) shocks are an accelerator of the CR particles up to knee energy ($\sim PeV$) still, very strong conclusive results are not present in support of this statement. Magnetic field disturbances in the proximity of the SNR shocks \citep{2004MNRAS.353..550B} is the main reason for the acceleration of the CRs in the well-accepted diffusive shock acceleration (DSA) model  \citep{1978ApJ...221L..29B, 1978MNRAS.182..147B}. According to the DSA model, the maximum energy of the acceleration of these CR particles is determined by their diffusion coefficient. The scattering of these CR particles is magnetic field disturbance dependent \citep{1966ApJ...146..480J}. Under turbulent cascade, these magnetic field disturbances possessing almost continuous energy spectrum resonates with the CR particles having different energy at different length scales concluding the fact that the energy spectrum is related to the maximum energy and diffusion coefficient of CR particles. If the energy spectrum of field disturbances follow the power-law i.e. $ \propto x^{m}$ then diffusion coefficient of the particles is also predicted to be a power law of the form $D(E) \propto E^{1-m}$, here $x$ and  $D(E)$ are length scale of the monochromatic field disturbances and diffusion coefficient of the particle with energy $E$ respectively  \citep{1987PhR...154....1B, 2006A&A...453..387P}. \cite{2006A&A...453..387P} have shown that for $m=2/3$  i.e. for Kolmogorov like \citep{1941DoSSR..30..301K} or trans-Alfvenic magnetohydrodynamic turbulence \citep{1995ApJ...438..763G}(GS95 henceforth), the maximum energy of the CR protons is very close to the knee energy under some shock compression models for the SNRs: Cassiopeia-A (Cas-A), Tycho and Kepler. For Cas-A i.e. for our case, the maximum energy of CR protons is almost consistent with the knee energy for the considered model of the shock compression ratio of 10. However observational evidence is required to establish such an important relationship. So it is important to determine the shape of the magnetic energy spectrum observationally in these SNRs and establish the relation between its shape and maximum energy of the CR particles (if single power law is found) as a conclusive statement. Here we determine to measure the shape of the magnetic energy spectrum from the synchrotron intensity in the SNR Cas-A. Our purpose in measuring the magnetic energy spectrum is to shed light on the few very important problems of the astrophysical aspect. First, if the 2/3 magnetic energy spectrum exists in nature? second, do Goldreich and Sridhar's (GS95) theoretical predictions of fully developed  Magnetohydrodynamic  turbulence for the trans-Alfvenic condition is consistent with observational 2/3 magnetic energy spectrum? and finally, if the slope of the magnetic energy spectrum and maximum energy of CR protons is specifically associated with $2/3$ law? Our results combined with earlier observational results of the magnetic energy spectrum in Tycho's SNR \citep{10.1093/mnras/sty2034} provides very important answers about the questions discussed above.\\
The shape of the magnetic energy spectrum can be determined using two-point correlation statistics of the synchrotron intensities and methodology is now well-established \citep{1959SvA.....3..415G,  doi:10.1080/10556799808232095, 2009MNRAS.393L..26R, 2010ApJ...720.1181C, 2012ApJ...747....5L,2016ApJ...818..178L, 10.1093/pasj/psx123}. However, these correlation analyses are biased since they include the geometry effect of the source in the estimator. Recently an unbiased method is discussed by \cite{10.1093/mnras/sty2034}  for sources having known spherical shape i.e. like spherical SNRs. We use this unbiased method to find the magnetic energy spectrum of the turbulent magnetic field in the SNR Cas-A.\\ 
In this work, we present our results for SNR Cas-A and discuss its
impact in the view of the astrophysical aspect. This work is structured in the following way.  In section~\ref{sec:method} we discuss briefly the methodology used in this work, followed by the analysis in section~\ref{sec:analysis}. We show the results of this analysis in section ~\ref{sec:results}. Finally, the importance of the results is discussed in section ~\ref{sec:discussion}.
\section{Method}
\label{sec:method}
The detailed method is presented in \cite{10.1093/mnras/sty2034}. Here we describe the methodology in very short. The two-point correlation function of the synchrotron intensity per frequency ($I_{\nu}$) used to  measure the magnetic energy spectrum  at radius r centred at the SNR's centre is given as
\begin{equation}
\begin{split}
C^{2}_{I_{\nu}}(\delta r) & =\frac{\int I_{\nu}(r)I_{\nu}(r')d^{2}r}{\int d^{2}r} \\
& \equiv  \langle I_{\nu}(r) I_{\nu}(r+\delta r) \rangle_{r}.
\end{split}
\label{eq:corr}
\end{equation}
where $r=r(x,y)$ is the two-dimensional sky position at radius r of the supernova and $r' =  r(x,y)+ \delta r(x,y)$. The notation r in the subscript is denoting that calculation  of the function $ C^{2}_{I_{\nu}}(\delta r)$ is done at radius $r(x,y)$ (with centre of supernova, say $x_{0},y_{0}$) having radial width  $\Delta r(x,y)$ such that $\Delta r <<r$. Note that $\delta r$ is the length scale of measurement while $\Delta r$ is the width of rim at radius r in which correlation function is measured. A very small value of $\Delta r $ in comparison to radius provides almost one-dimensional measure of  $C^{2}_{I_{\nu}}(\delta r)$.   The requirement of the spherical supernova is to avoid any effect of the geometry in the measurement of second-order correlation function. Calculation of the $C^{2}_{I_{\nu}}(\delta r) $ at radius r from centre in a spherical object guarantees that the depth along z-axis at radius r  is constant, making  $C^{2}_{I_{\nu}}(\delta r) $ free from geometry effect. To clarify the fluctuating component of the magnetic field we use correlation statistics $|C^{2}_{I_{\nu}}(\delta r)-C^{2}_{I_{\nu}}(\delta r_{min})|$. Here $\delta r_{min}$ represents the minimum value of $\delta r$. If the correlation statistics follows  power law then 
\begin{equation}
|C^{2}_{I_{\nu}}(\delta r)-C^{2}_{I_{\nu}}(\delta r_{min})|= A \delta r^{\alpha}
\label{eq:stat}
\end{equation}
Where A and $\alpha$ are the amplitude and power-law index of the power-law. From the theory of error propagation, it is  known that uncertainty associated with function $|C^{2}_{I_{\nu}}(\delta r)-C^{2}_{I_{\nu}}(\delta r_{min})|=A \delta r^{\alpha}$ is related to the uncertainty measure in  $\delta r$ as   
\begin{equation}
\sigma_{C}=A \alpha \delta r^{\alpha-1} \sigma_{\delta r}
\label{eq:error}
\end{equation}
 Here $\sigma_{\delta r}$ is the uncertainty in the measurement of $\delta r$ while $\sigma_{C}$ represents the uncertainty associated with function $|C^{2}_{I_{\nu}}(\delta r)-C^{2}_{I_{\nu}}(\delta r_{min})|$ for the power-law case. We use equation~\ref{eq:error} to calculate the uncertainty $\sigma_{C}$ from known uncertainty measurement in $\delta r$ for the cases where the magnetic energy spectrum follows the power-law. To quantify the magnitude of the fluctuations we plot the normalized statistics $|C^{2}_{I_{\nu}}(\delta r)/C^{2}_{I_{\nu}}(0) -1|$. In the case, statistics of the second-order correlation function of the magnetic energy is of trans-Alfvenic nature, $\alpha$ must be 2/3 within measurement uncertainty.  
\section{Analysis}
\label{sec:analysis}
\begin{figure}
\hspace*{-0.25cm}
\includegraphics[scale=0.55]{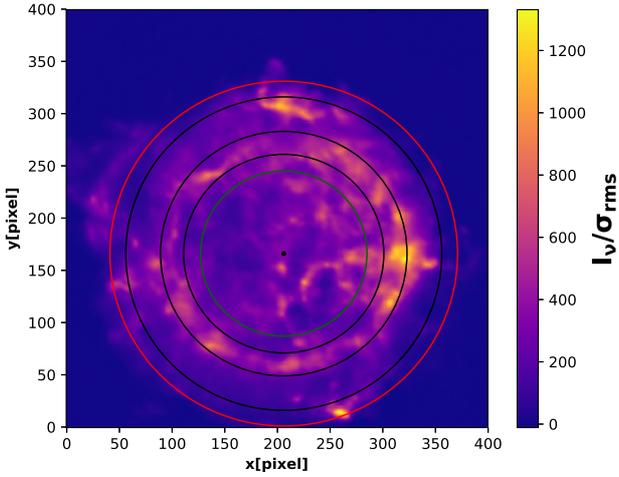}
\caption{Radio continuum image of the Cassiopeia-A (normalized by rms). Here x and y axes are in the units of pixels and 1 pixel has a size of 1". The dark-green radius corresponds to 0.520R(78", innermost circle), red corresponds to 1.093R(164", outermost circle) and 3 consecutive  black radius corresponding to 0.633R(95"), 0.780R(117") and 1.000R(150") show mean radius of the regions where trans-Alfvenic magnetohydrodynamic turbulence is found (as shown in  figure~\ref{fig:figure2} and \ref{fig:figure4}). The investigated radial region is from 0.520R to 1.093R (dark-green to red).}
\label{fig:figure1}
\end{figure}

 Supernova Cas-A is known to be a spherical shell in shape \citep{1995ApJ...440..706R} with radius (R) of $\sim 2.5'$. Here we analyze the continuum image of the Cas-A observed using  GMRT \citep{1991CuSc...60...95S} interferometer in radio band 410-460MHz and published in \cite{2019MNRAS.486...42C}. The image has a pixel size of the 1". Details of the observation and data analysis are well presented in \cite{2019MNRAS.486...42C}. The RMS normalized image is shown in figure~\ref{fig:figure1}. To measure the magnetic energy spectrum via second-order autocorrelation function (equation~\ref{eq:corr}) of synchrotron intensities at different radius r of SNR, we have chosen centre of Cas-A at  23:23:26.10,  +58:48:53.70(J2000). Centre chosen above is almost the same as described radio centre of Cas-A at 23:23:26,  +58:48:54(J2000) by \cite{2018A&A...612A.110A}. This is very close to the reverse shock centre at 23:23:25.44, +58:48:52.3 (J2000) (see \cite{2001ApJ...552L..39G}). To measure $C^{2}_{I_{\nu}}(\delta r)$ we choose circles of 3-pixel width at all radius of interest from the centre of the SNR. This circle width corresponds to the 3" in size. Chosen width is almost negligible in comparison to the radius R (150") of the supernova.  We use pixels having $I_{\nu}> 3 \sigma_{rms}$ from the image to avoid any type of noise statistics in our estimator. Calculation of the two-point correlation function (defined in equation~\ref{eq:corr}) is done from radius 0.526R (78") to 1.093R (164") in step of radial increment $\Delta r\sim 0.006$R (1"). In the above region of investigation, we found that statistics $|C^{2}_{I_{\nu}}(\delta r)-C^{2}_{I_{\nu}}(\delta r_{min})|$  can be represented as a power law in the following 3 radial regions, from  0.613R-0.653R (region 1), 0.760R-0.800R (region 2) and 0.980R-1.020R (region 3). At the smallest scale, we found that measured fluctuation is not very right on the power-law trend because of the bright knots in the SNR that give excess power at the lowest scale. These knots also make SNR Cas-A little bit deviated from the spherical shape but the possible effect of this on measured spectra is expected to be negligible as the deviation from the spherical shell itself is small in comparison to its radius and Cas-A is known to be almost of spherical shell in shape (see \citep{1995ApJ...440..706R}, \cite{2018A&A...612A.110A} and other references available in this text). Errors in the $|C^{2}_{I_{\nu}}(\delta r)-C^{2}_{I_{\nu}}(\delta r_{min})|$ are calculated using equation~\ref{eq:error}. We take $\sigma_{\delta r}$ as an uncertainty measurement in $\delta r$ to calculate uncertainty $\sigma_{C}$ in $|C^{2}_{I_{\nu}}(\delta r)-C^{2}_{I_{\nu}}(\delta r_{min})|$ for the power-law cases. To find A and $\alpha$ properly, first we fit the measured spectrum assuming power law of  2/3 to calculate values of A and $\alpha$ as a zeroth-order approximation, then we use these fitted values of  A and $\alpha$ in equation~\ref{eq:error} to calculate $\sigma_{C}$. Now using this $\sigma_{C}$ we again fit $|C^{2}_{I_{\nu}}(\delta r)-C^{2}_{I_{\nu}}(\delta r_{min})|$  to get values of  A and $\alpha$ as a first-order approximation. We repeat this process until we find that there is no further improvement in the values of A and $\alpha$. In our analysis, it is found that fitted values of A and $\alpha$ in zeroth-order approximation and higher-order approximations are almost the same and do not differ significantly. Even values obtained in the first-order approximation are sufficiently stable. We also calculate $C^{2}_{I_{\nu}}(0)$ and found that $C^{2}_{I_{\nu}}(\delta r_{min}) \sim C^{2}_{I_{\nu}}(0)$. Since field disturbances at smaller scales are guided by field disturbances at larger scales \citep{2000ApJ...539..273C}, so the maximum value of $\delta r$ is taken such that  $|C^{2}_{I_{\nu}}(\delta r)/C^{2}_{I_{\nu}}(0) -1|$  is at least less than 0.5 (in fact it should be even smaller). The larger value of the $|C^{2}_{I_{\nu}}(\delta r)/C^{2}_{I_{\nu}}(0) -1|$ does not represent the true magnetic energy spectrum as assumption that $I_{\nu}$ scale linearly with the fluctuations of the magnetic field, remains no longer valid. For region 1, fitted range of length scale is 8"-65". Since there is a break in amplitude at $\delta r \sim $ 80" for region 1 so we do not fit beyond this length scale but compare the spectra with 2/3 law above 80". To show that even beyond 80" spectra are still close to 2/3 law, we fit it in the range of 85"-170" at radius 0.647R and show in  figure~\ref{fig:figure3} (top panel). Spectra for the regions 2 and 3 are fitted in the range of 9"-182" and 10"-105" respectively. So through many iterations of the fitting, we could define the "good range of fit" for SSR, RSR and FSR as 8"-65", 9"-182" and 10"-105"
respectively. Beyond these limits fitting parameters are not well compatible and fitted power law is not in good agreement with the measured magnetic spectrum.

\section{Results}
\label{sec:results}
\begin{figure}
\includegraphics[scale=0.52]{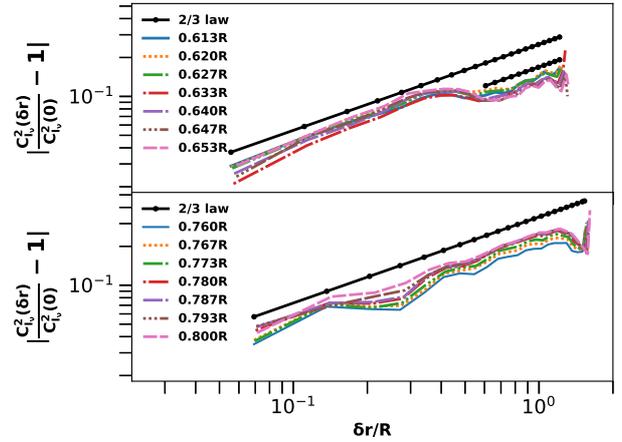}
\caption{Plots of the 7 best magnetic energy spectra of the two regions (region 1 and region 2)  i.e. from 0.613R to 0.653R and from 0.760R to 0.800R where power law is fitted over the decade of length scales. Table  ~\ref{tab:table1} lists values of $\alpha$ with 1 $\sigma$ error bars for the above spectra. The x-axis shows the length scales in unit of radius R (150") of supernova. We show the fluctuating component of the correlation,  $|C^{2}_{I_{\nu}}(\delta r)/C^{2}_{I_{\nu}}(0) -1|$ on y-axis.}
\label{fig:figure2}
\end{figure}

\begin{figure}
\includegraphics[scale=0.52]{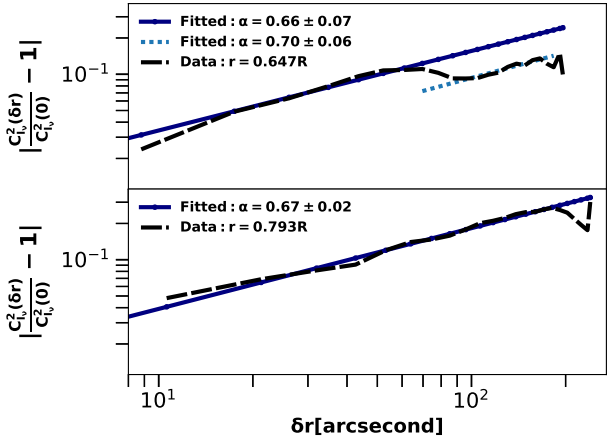}
\caption{The two best fitted magnetic energy spectra from  figure~\ref{fig:figure2}. Here we show the best fit and data at radius 0.647R and 0.793R. The x-axis is in the arc-second unit and y-axis is same as in  figure~\ref{fig:figure2}.}
\label{fig:figure3}
\end{figure}

\begin{figure}
\includegraphics[scale=0.52]{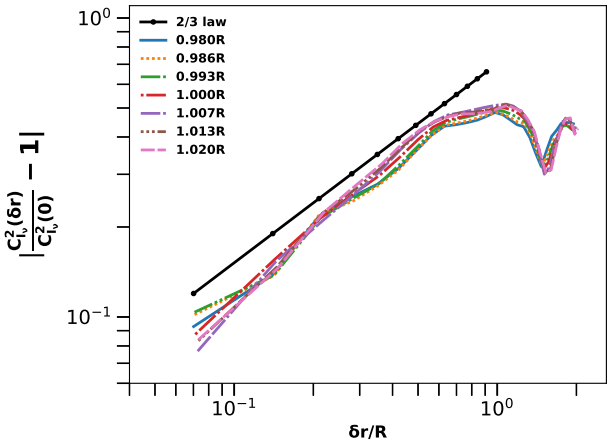}
\caption{The seven best chosen magnetic energy spectra from 0.980R to 1.020R (region 3). Fitted values of $\alpha$ are listed in table  ~\ref{tab:table1}. The x and y axes are same as in  figure~\ref{fig:figure2}.}
\label{fig:figure4}
\end{figure}

\begin{figure}
\includegraphics[scale=0.52]{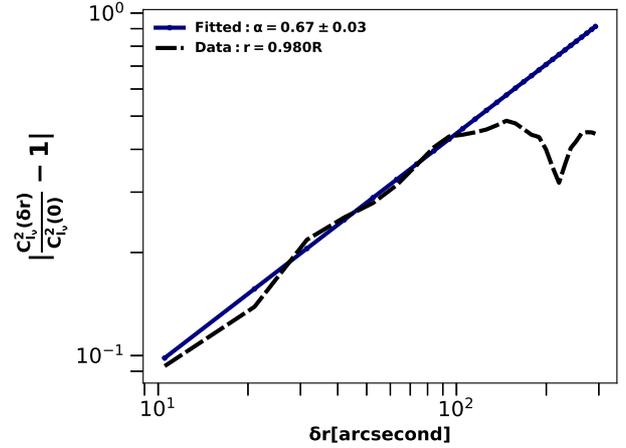}
\caption{One of the best-fitted spectrum from the  figure~\ref{fig:figure4}. Shown x and y-axes are the same as in  figure~\ref{fig:figure3}.}
\label{fig:figure5}
\end{figure}

 Figure~\ref{fig:figure2} shows the best measured magnetic energy spectra from region 1 and region 2  where statistics of the fluctuation have power-law form. We show combine results in one figure from two regions since they belong to the near reverse shock position. The top panel in  figure~\ref{fig:figure2} is from region 1 while the bottom panel shows the spectra from region 2. We show normalized statistics  $|C^{2}_{I_{\nu}}(\delta r)/C^{2}_{I_{\nu}}(0) -1|$ on y-axis and length scale in terms of radius R of supernova on x-axis. In each panel,  seven correlation functions (magnetic energy spectra) from each above regions are shown. The selection of these correlation functions from each region is based on the best reduced $\chi^{2}$ and the power-law index  $\alpha \simeq 2/3$ within measurement uncertainty. For comparison, 2/3 law is plotted point to point in each panel of  figure~\ref{fig:figure2}. Since the amplitude (normalized) of each spectrum is nearly the same and difference in slope is almost insignificant, so to properly differentiate the spectrum we arbitrarily scale the amplitude, make it index free and show in figure \ref{fig:figure9}.  Figure~\ref{fig:figure3} shows two best-chosen correlation functions from region 1 (top figure)  and region 2 (bottom figure) with their fitted values of the $\alpha$ shown in the top left corners. These chosen correlation functions in figure \ref{fig:figure3} are such that their measured power-law index ($\alpha$) is closest to 2/3 than the other 6 correlation functions in the same region. Y-axis in figure~\ref{fig:figure3} is same as in  the figure~\ref{fig:figure2} while x-axis is shown in unit of arc-second. In  figure~\ref{fig:figure4} we show the spectra for region 3 in similar way as shown in  figure~\ref{fig:figure2} and discriminate individual spectrum in figure \ref{fig:figure10}. Fitted values of the $\alpha$ with 1 $\sigma$ error bars for the spectra shown in figure \ref{fig:figure2} and figure~\ref{fig:figure4} are tabulated in table~\ref{tab:table1}. From table~\ref{tab:table1}  it is also clear that all fitted values of  $\alpha $ are 2/3 within 1.5$\sigma$ error bars. Similar to the figure \ref{fig:figure3},  in figure \ref{fig:figure5} we show one of the best chosen energy spectrum from figure \ref{fig:figure4}. Energy  spectra  investigated at all radius (from 0.526R"-1.093R") are systematically shown in the appendix (figures ~\ref{fig:figure11} and \ref{fig:figure12}). The plotted value of the dimensionless spectrum $|C^{2}_{I_{\nu}}(\delta r)/C^{2}_{I_{\nu}}(0) -1|$ at each radius in figures ~\ref{fig:figure11} and \ref{fig:figure12} makes it easier to identify the maximum value of $\delta r$. The range of the fitted spectra for the regions 1, 2 and 3 are also marked in the same figures (\ref{fig:figure11} and \ref{fig:figure12}) with the dashed vertical lines.

\begin{center}
\begin{table}
\begin{tabular}{ |p{0.9cm}|p{0.9cm}|p{0.9cm}|p{0.9cm}|p{0.9cm}|p{0.9cm}| }
 \multicolumn{6}{|c|}{ }\\
 \hline
 radius(r/R)& \quad$ \hspace{0.22cm} \alpha$ & radius(r/R) & $ \quad \quad  \hspace{-0.05cm} \alpha$ & radius(r/R) & $\quad \quad \hspace{-0.05cm} \alpha$ \\
 \hline
 0.613 & 0.64$\pm$0.02 & \hspace{0.2cm}  0.760 &  0.66$\pm$0.03& \hspace{0.2cm}  0.980& 0.67$\pm$0.03\\
 0.620 & 0.68$\pm$0.03 & \hspace{0.2cm}  0.767 &  0.68$\pm$0.03& \hspace{0.2cm}  0.986& 0.68$\pm$0.03\\
 0.627 & 0.67$\pm$0.03 & \hspace{0.2cm}  0.773 &  0.70$\pm$0.03& \hspace{0.2cm}  0.993& 0.70$\pm$0.03\\
 0.633 & 0.72$\pm$0.04 & \hspace{0.2cm}  0.780 &  0.69$\pm$0.03& \hspace{0.2cm}  1.000& 0.67$\pm$0.03\\
 0.640 & 0.70$\pm$0.06& \hspace{0.2cm}  0.787 &  0.69$\pm$0.03& \hspace{0.2cm}  1.007& 0.70$\pm$0.04\\
 0.647 & 0.66$\pm$0.07& \hspace{0.2cm}  0.793 &  0.67$\pm$0.02& \hspace{0.2cm}  1.013& 0.67$\pm$0.04\\
 0.653 & 0.63$\pm$0.07& \hspace{0.2cm}  0.800 &  0.64$\pm$0.02& \hspace{0.2cm}  1.020& 0.64$\pm$0.05\\
 \hline
\end{tabular}
\caption{Fitted values of $\alpha$ for the spectra shown in  figures~\ref{fig:figure2} and \ref{fig:figure4}}
\label{tab:table1}
\end{table}
\end{center}

\section{Discussion}
\label{sec:discussion}
The correlation functions for the regions where 2/3 law holds i.e. for region 1,  region 2 and region 3 shows well developed trans-Alfvenic turbulence as described by \cite{1995ApJ...438..763G}. If we take Cas-A distance 3.4kpc \citep{1995ApJ...440..706R} and radius $2.5'$ then the radial window  of the well developed turbulence with $\alpha \simeq 2/3$ has width of $\sim$0.047R (0.115pc). GS95 shows that such structures following the 2/3 power law are a reflection of the fully developed MHD turbulence, equivalent to the Alfven Mach no. $M_{A} \simeq 1$ ( for detail discussion see \cite{Lazarian_2016}). In this case  turbulence at the injection scale L have nearly same injection velocity $u_{i}$ and Alfven velocity $V_{A}$ such that $M_{A}=u_{i}/V_{A}\simeq1 $. We will discuss shortly how this condition is satisfied for the regions where 2/3 magnetic energy spectra are observed in the SNR Cas-A. In the dynamical evolution of the SNR, a contact discontinuity exists between the reverse and forward shocks \citep{Laming_2001}. Radio reverse shock radius described by  \cite{2018A&A...612A.110A}  for the SNR Cas-A is at $0.760\pm0.040$R (114"$\pm$6"). Note that, region 2 where spectra with $2/3$ law are found is in the vicinity of this reverse shock position so we call region 2 as reverse shock region (hence-after RSR). Cas-A reverse shock radius using X-ray data is found to be at 0.633R$\pm 0.066$ (95"$\pm$10") \citep{2001ApJ...552L..39G} (with the almost same centre) shows that region 1, where $2/3$ law again holds is in the tight contact with this X-ray reverse shock position so we call region 1 as sub shock region (hence-after SSR). In \cite{2018A&A...612A.110A}, authors noticed that the reverse shock observed in the radio and X-ray frequency observations differ from each other and have different radii. The radio reverse shock is more centrally located (in comparison to expansion centre) than the X-ray reverse shock and they coincide at the western region of the SNR. The most probable reason that the X-ray reverse shock differs from the radio is that the X-ray reverse shock definition is based on the observation of the X-ray synchrotron emitting filament. In the view of the facts that our measured magnetic energy spectra of 2/3 power law at SSR is in the close vicinity of the X-ray reverse shock position and such type of fluctuations are developed in the vicinity of the shocks (discussed below), the spectra at SSR are most likely the reflection of the observed X-ray reverse shock amplifying the magnetic field at this location. We refer this radial position as a sub-shock region (SSR) because measured spectra (in the radio frequency observation) at this location have cut-off and are not fully developed to be an indicator of a shock rather than possessing the property of a semi-shock. In the same work of X-ray data forward shock for Cas-A is found at radius  1.020R$\pm0.080$R(153"$\pm$12") \citep{2001ApJ...552L..39G}. This shows that region 3 with the energy spectra of 2/3 law is in the vicinity of forward shock position, so we call region 3 forward shock region (hence-after FSR).  Contact discontinuity in SNR Cas-A is found to exist between these forward and reverse shocks \citep{Laming_2001}. From the above discussion, it is clear that well developed MHD turbulence with $\alpha \simeq$2/3 is in the vicinity of the SNR shocks. The condition of satisfying $M_{A} \simeq 1$ with $\alpha \simeq 2/3$ in the vicinity of the shock can be explained by the amplification of the magnetic field and it is well discussed by \cite{10.1093/mnras/sty2034} owing to the widely accepted numerical simulations and observational references. \cite{2004A&A...419L..27B} have confirmed the magnetic field amplification in the SNR Cas-A from Chandra X-ray data by providing direct evidence of CR acceleration. \\
Here we verify our results for the trans-Alfvenic like scaling, considering an amplified magnetic field in the vicinity of the shocks. For comparison, if we take shock speed as injection velocity at largest scale L and downstream amplified magnetic field of Cas-A $\sim$ 485$\mu$G \citep{2005A&A...433..229V} in the proximity of the reverse shock with shock speed ~2400 km/s,  we get Alfven Mach number $\sim$ 1.8 (considering the ambient density of  Cas-A $\sim 10^{-24} g cm^{-3}$ \citep{2010PNAS..107.7141B}). Using Doppler mapping technique on X-ray data \cite{2002A&A...381.1039W} found forward shock speed $4000\pm 500 $ km/s at radius r = 1.020R (153") for SNR Cas-A. This provides an upper limit of $M_{A} \simeq 3.3$ for SNR Cas-A. Note that these are global estimates which are almost consistent with the GS95 trans-Alfvenic prediction. To get the exact value of the Alfven Mach no. at these positions with the measured magnetic spectra of 2/3 law, we need a very precise value of the magnetic field, shock speed, and ambient density at these radial positions. Unfortunately, such a census with very good precision is not available. If we consider the results of the magnetic energy spectra in Tycho's SNR as discussed by \cite{10.1093/mnras/sty2034} with the shock speed  $\sim$ 5000km/s \citep{2016ApJ...823L..32W} and amplified downstream magnetic field of Tycho's SNR as $\sim$ 273$\mu G$ \citep{2005A&A...433..229V}, we get Alfven Mach no. $\sim$ 1.8 which is again globally consistent with our expectations of the trans-Alfvenic MHD turbulence condition. Our results along with Tycho's, show 2/3 magnetic energy spectra developed in the vicinity of the SNR shocks of Cas-A and Tycho. These two results with the verified global Alfven Mach no. confirm the 2/3 magnetic energy spectra in these two young SNRs. Found 2/3 law (in Cas-A) at SSR predicts a sub shock existing at this radial position and contact discontinuity between SSR and RSR. This suggests that SNR Cas-A shows an additional subshock in radio frequency observations along with forward and reverse shocks. So almost over the radial width of 0.047R, 2/3 law is consistent for the given three regions i.e. for SSR, RSR and FSR  implying the well amplification of magnetic field over the radial window of $\sim 0.115$pc. Beyond this width, turbulence is not well developed as a trans-Alfvenic. There are known instabilities  like Richtmyer-Meshkov instability (RMI) ( see \cite{doi:10.1002/cpa.3160130207}, \cite{Sano_2012} and \cite{2013ApJ...772L..20I}), acoustic instability \citep{1986MNRAS.223..353D} and Bell instability \citep{2004MNRAS.353..550B} that exist near the shock and may amplify the magnetic field in the vicinity of SNR shocks. To explore the exact mechanism of magnetic field amplification near these SNR shocks, a subarcsecond resolution is required and is the goal of the future science with Square Kilometre Array. Bell instability results in upstream magnetic field amplification near the shock and may cause a trans-Alfvenic situation. Our results for RSR  showing almost single power-law over a decade of range scales maybe because of the already amplified magnetic field due to Bell's instability at this position. Energy spectra for the SSR having a break in amplitude $\sim$ 80" shows either evolutionary stage of GS95 turbulence or strong interaction between GS95 turbulence and other types of instabilities (mentioned earlier ) at this scale. For the FSR region, beyond the length scale of 110"  spectra deviates from the 2/3 law and shows somewhat flatter characteristics. Similar behavior is observed in the Tycho's results for the spectra below the radius of 0.85R. This is supposed to be the interaction between RMI instability and GS95 turbulence \citep{10.1093/mnras/sty2034}. The shape of the power-law of the magnetic energy spectra from these two SNRs is found to be single power-law over the decade of the length scales as predicted by (GS95,\cite{2000ApJ...539..273C}). However, more results are required to support the prediction of  "single power-law " magnetic energy spectra in the SNRs as a general statement. The results of possible investigations will be presented in the near future. As from Cas-A and Tycho's results, it is evident that the magnetic energy spectra in these SNRs is of trans-Alfvenic nature, so they should have a maximum energy of the CR particles reaching near to the knee energy under the model discussed in \citep{2006A&A...453..387P}. But Cas-A is the TeV candidate not the PeV at the present stage \citep{Zhang_2019}. This shows that the maximum energy of the CR protons needs to be constraint with different shock compression ratio models rather than 4 and 10 as discussed by \cite{2006A&A...453..387P} for the present stage. \\
To summarise, the magnetic energy spectra in the SNRs Cas-A and Tycho near the shocks are 2/3 (Kolmogorov like) over the decade of length scales and show developed trans-Alfvenic magnetohydrodynamic turbulence in these two young  SNRs. Magnetic energy spectra in SNR Cas-A suggest it to be investigated as two shock supernova along with a sub shock and need to be studied with a plausible shock compression ratio models rather than discussed in  \cite{2006A&A...453..387P} to explain the observed maximum energy of CR protons with 2/3 energy spectrum at the current stage.
\section*{ACKNOWLEDGMENT}
We are very much thankful to Aditya Chowdhury for providing us the GMRT image of the Cassiopeia-A to carry out this research work. GMRT is the national facility for radio astronomical research at meter-wavelength run by NCRA, India. We would also like to thank Sanjeet Kumar Patel and Aditya Chowdhury for many useful discussions during this work. PKV would like to thank MHRD India for financial support during this work through senior research fellowship (SRF). JK acknowledges the University Grants Commission, India for providing financial support through the Senior Research Fellowship. Authors thank the anonymous referee for their concise and constructive suggestions that have improved the presentations of the paper significantly. 
\section*{DATA AVAILABILITY}
The data used for the analysis and findings within this paper will be shared by the corresponding author upon reasonable request.        
\bibliographystyle{mnras}
\bibliography{casaref} 
\appendix

\section{ }
\label{sec:append}
 Figures ~\ref{fig:figure11} and \ref{fig:figure12} show the all measured magnetic energy spectra from 0.526R (dark-green circle in  figure~\ref{fig:figure1}) to 1.093R (red circle in  figure~\ref{fig:figure1}) that have been investigated from the radio centre of supernova Cas-A (Color figure is available in the online version of this work). The investigated spectra are assumed to be isotropic in nature as widely studied for continuum power spectrum as well as to extract HI optical depth spectra of ISM using continuum power spectrum (see \cite{10.1111/j.1745-3933.2010.00831.x}, \cite{10.1093/mnras/stz3148} and \cite{2018ARA&A..56..489S}). The whole range is divided into two figures having 4 sub-figures in each. The division of the figures is done to clarify : 1) how the transition of the spectrum takes place near the shock positions and  2) regions beyond the shock positions do not show power-law like statistics.
\begin{figure*}
\hspace*{-0.6cm}
\includegraphics[width=13.5cm,height=10.5cm]{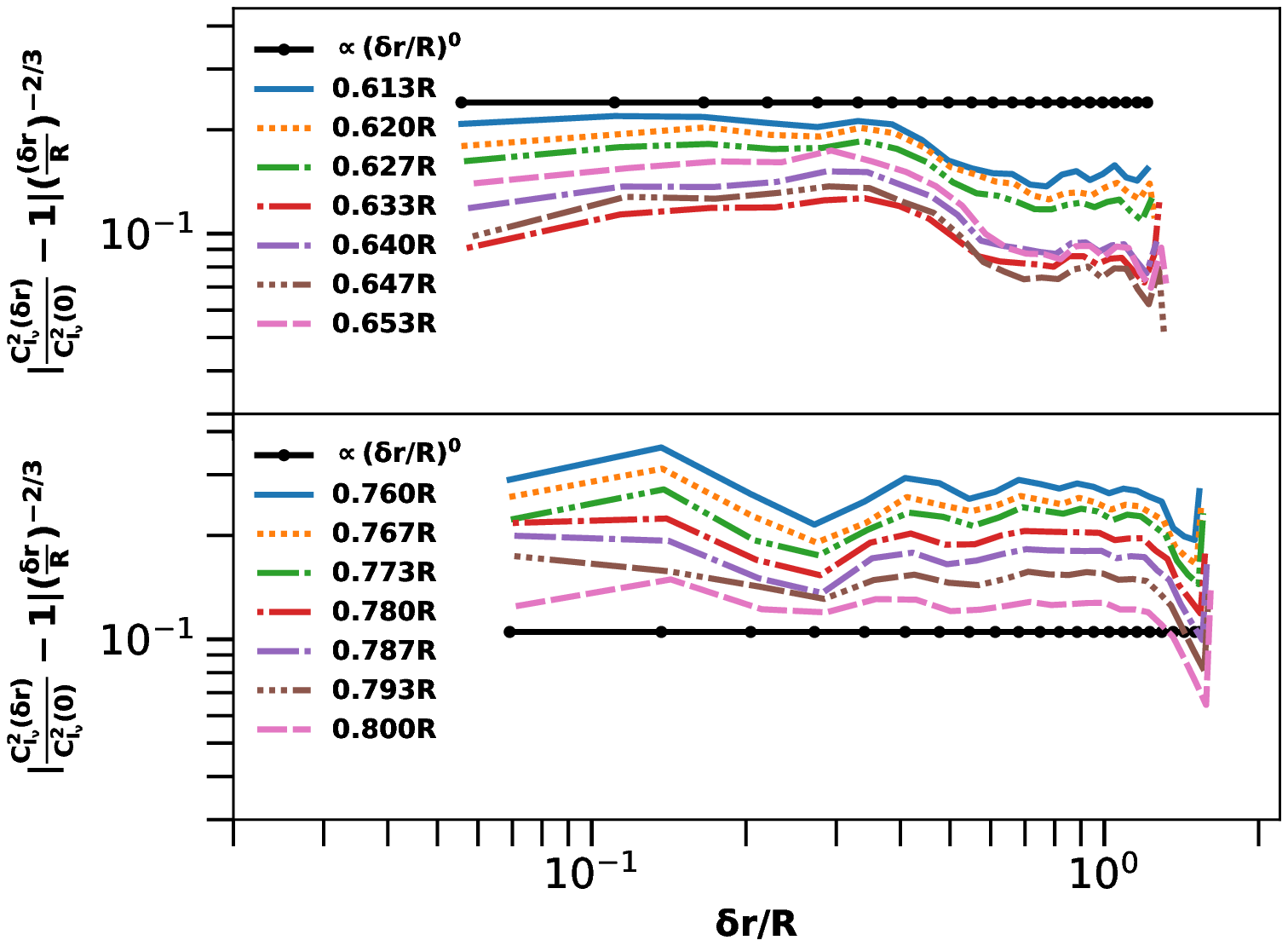}
\caption{The above figure shows the modified version of figure ~\ref{fig:figure2}. To differentiate individual spectrum, we make it index-free ( $(\delta r/R)^{0}$) by multiplying $(\delta r/R)^{-2/3}$ and scaling the amplitude arbitrarily at each radius.} 
\label{fig:figure9}
\end{figure*}

\begin{figure*}
\hspace*{-0.6cm}
\includegraphics[width=13.5cm,height=8.0cm]{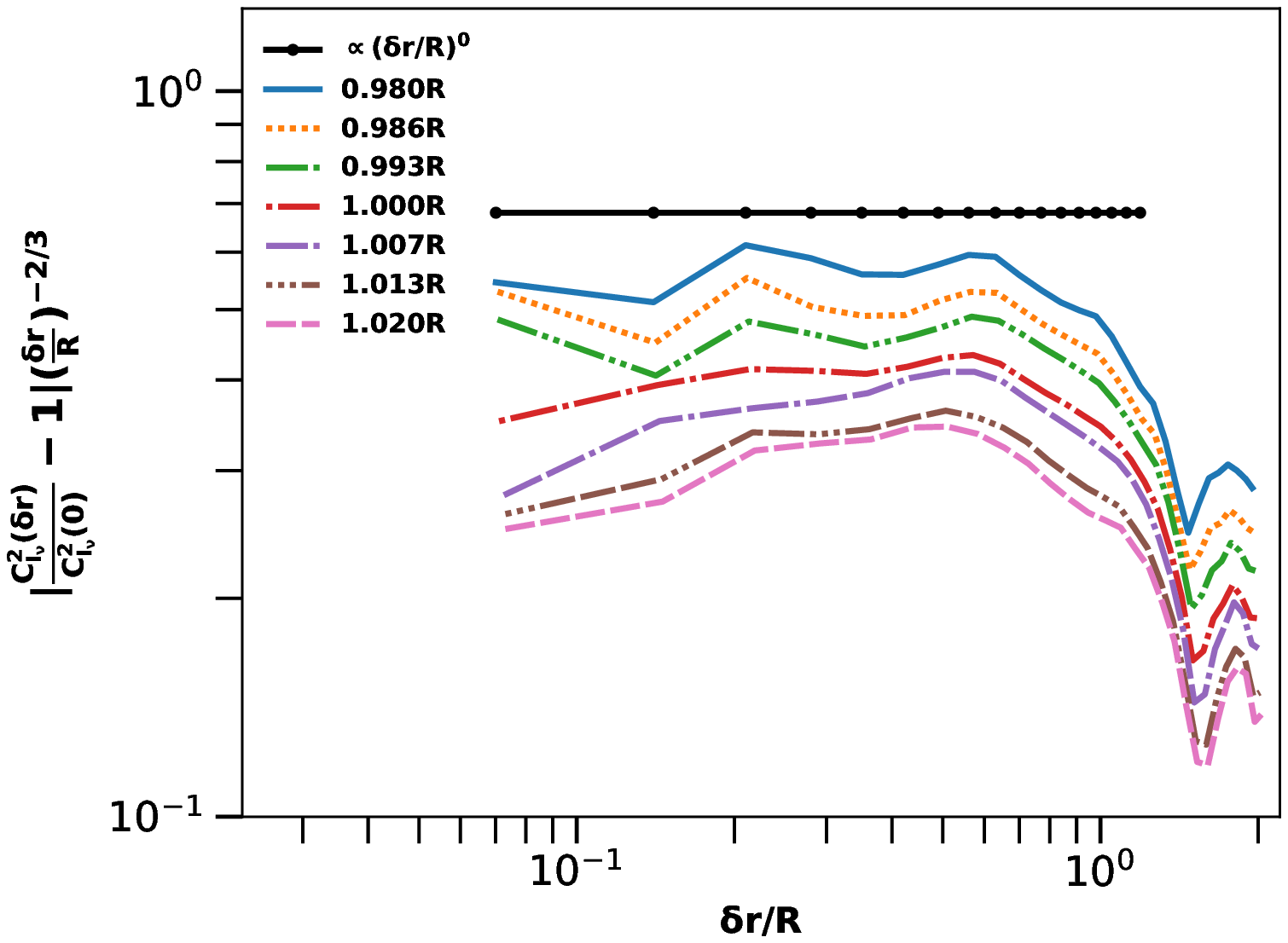}
\caption{The modified version of figure ~\ref{fig:figure4} and it is similar plot as figure ~\ref{fig:figure9}.}
\label{fig:figure10}
\end{figure*}

\begin{figure*}
\hspace*{-0.1cm}
\includegraphics[width=15cm,height=10.0cm]{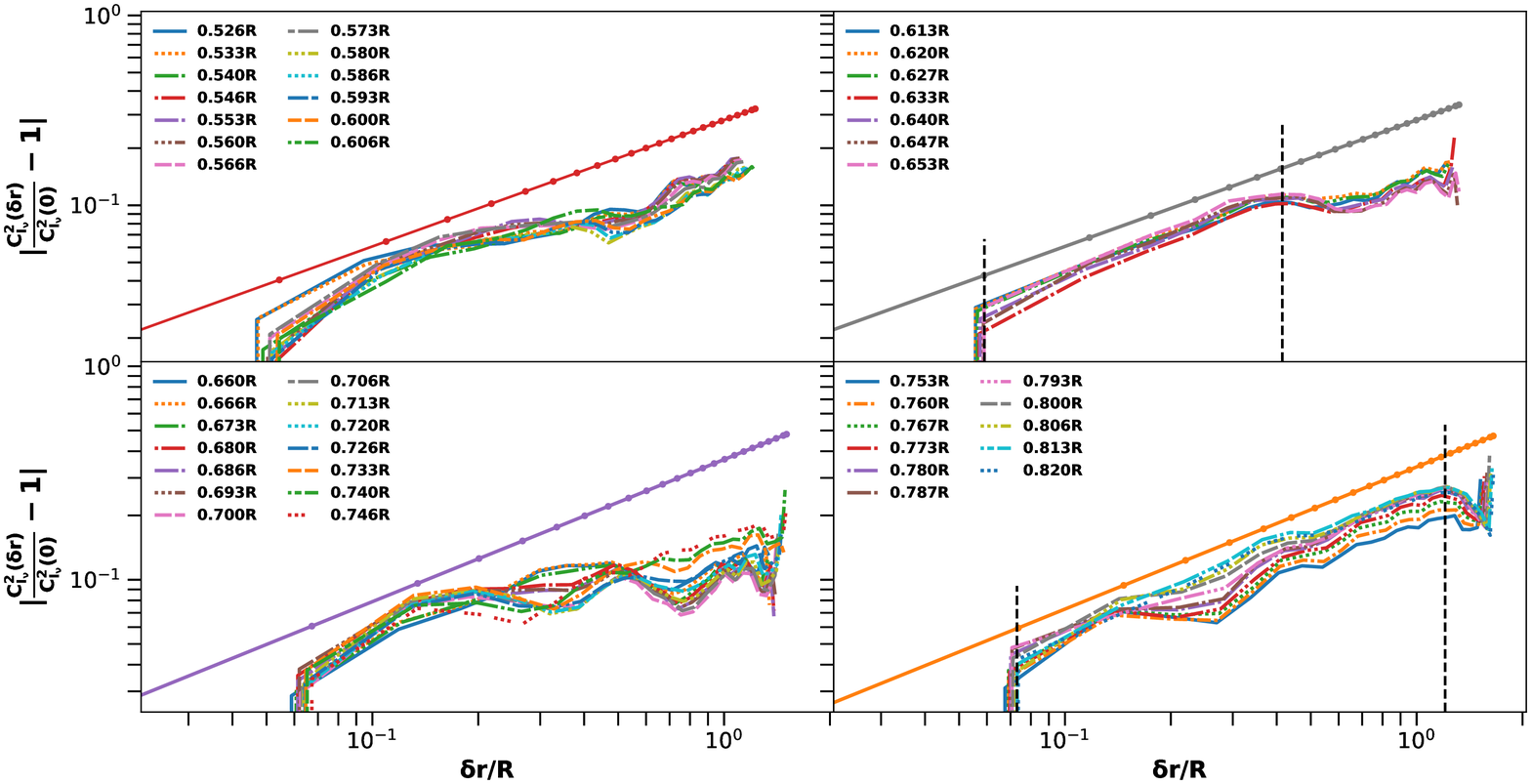}
\caption{The figure shows the normalized magnetic energy spectra from 0.526R to 0.820R divided into four groups. The straight line with dot points corresponds point to point 2/3 law and is plotted for comparison. Vertical dotted lines in subfigures (top right and bottom right) shows the fitted range of length scales. From all four regions, we see how the magnetic energy spectra evolves radially and transition takes as a function of the radius near the shocks. It is clear from the figure that in the regions from 0.613R-0.653R and 0.753R-0.820R the energy spectra have a power-law form. We have chosen the best 7 spectra from the regions 0.613R-0.653R and 0.753R-0.820R to show in  figure~\ref{fig:figure2} and tabulate their fitted values of $\alpha$ in table~\ref{tab:table1}.}
\label{fig:figure11}
\end{figure*}

\begin{figure*}
\includegraphics[width=15cm,height=10.0cm]{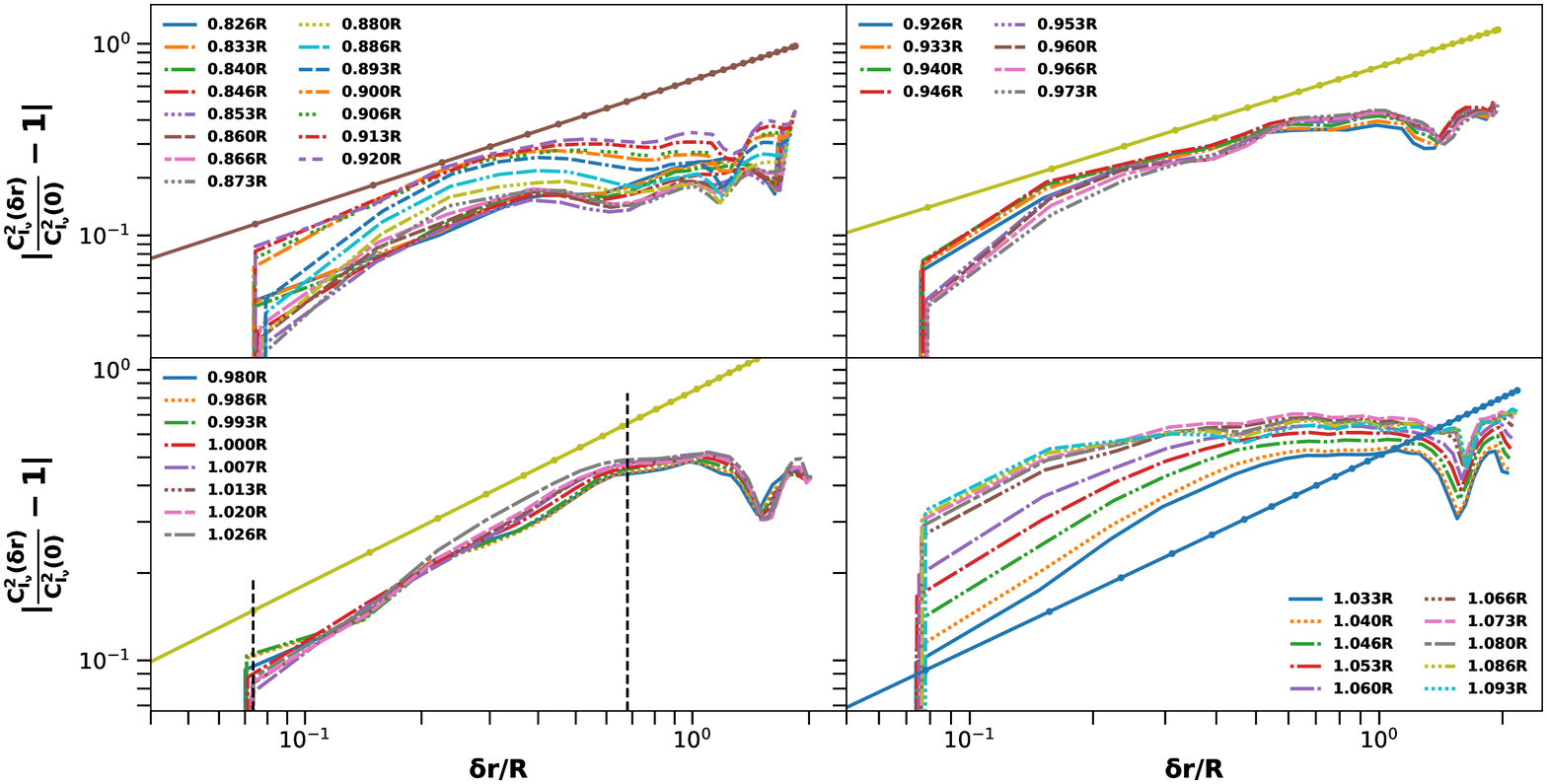}
\caption{Plots of the normalized magnetic energy spectra from 0.826R to 1.093R showing that only the region 0.980R-1.026 has the power-law statistics over considerable length scales. Vertical dotted lines in subfigure (bottom left) shows the fitted range of length scales for region 3. The 7 best-chosen spectra from this range is shown in  figure~\ref{fig:figure4}  and their fitted values the $\alpha$ is tabulated in  table~\ref{tab:table1}.}
\label{fig:figure12}
\end{figure*}

\bsp	
\label{lastpage}
\end{document}